\documentclass[10pt,a4paper]{article}
\usepackage{amsmath, amsfonts, amssymb}
\usepackage{graphicx}
\usepackage{url}
\usepackage{natbib}

\begin{document}
\title{\bf \large Shadows around  Sgr A* and  M87* as a tool to test gravity theories}
\author{Alexander F.~Zakharov\thanks{E-mail: alex.fed.zakharov@gmail.com}
}
\date{\it  \small   Bogoliubov Laboratory for Theoretical Physics, JINR,
141980 Dubna, Russia, \\
}

\maketitle

\begin{abstract}
In the framework of Randall -- Sundrum theory with extra dimension Reissner -- Nordstr\"om black hole solutions with a tidal charge have been found. The shadow around the supermassive black hole in M87 was reconstructed in 2019 based on observations with the Event Horizon Telescope (EHT) in April 2017. In May 2022 the EHT Collaboration presented results of a shadow reconstruction for our Galactic Center.
Earlier, for Reissner -- Nordstr\"om metric we derived analytical expressions for shadow size as a function of charge and later generalized these results for a tidal charge case.
We discuss opportunities to evaluate parameters of alternative theories of gravity with shadow size estimates done by the EHT Collaboration, in particular, a tidal charge could be estimated  from these observations.

\end{abstract}

{\it Key words}: Supermassive black holes, Galactic Center, M87, Synchrotron radiation, VLBI observations.

\section{\large Introduction}

Several years ago the Event Horizon Telescope (EHT) Collaboration  has been formed. The astronomers use telescopes located over the globe and they are operating at 1.3~mm wavelenth in VLBI regime or in other words the corresponding network acts as a giant telescope with Earth size.
An angular resolution of the network is around 25~$\mu as$ which is comparable with angular sizes of event horizons for supermassive black holes in
Sgr A* and M87*. In spite of huge differences in black hole masses and distances toward these objects these shadows have similar sizes (52~$\mu as$ for Sgr A* and 42~$\mu as$ for M87*) In April 2017 the EHT Collaboration observed the Galactic Center and the center of M87 galaxy. In 2019  the EHT Collaboration presented results of a shadow reconstruction for M87* and in May 2022 a shadow reconstruction  for Sgr A* was presented by the EHT Collaboration. Now these images with shadows around M87* and Sgr A* are used as logos for these supermassive black holes or sometimes more generally for any astrophysical black holes. Since it is impossible
to observe dark regions (shadows) astronomers found bright regions of synchrotron emission in 1.3~mm and reconstructed shapes and sizes of shadows.
These remarkable achievements in precise observations and data analysis are based on three pillars: synchrotron emission which is generating in many astronomical objects including environments of supermassive black holes, VLBI ideas which were efficiently implemented in the EHT network and relativistic analysis of geodesics in the black hole metrics.

\section{\large Synchrotron radiation}

Radiation of electrons moving in magnetic fields  (it is now called synchrotron one) was discussed in details in a fundamental book by  \cite{Schott_12} but at these times there were no facilities to detect it in experiments or in astronomical observations\footnote{In recent book \cite{Connerade_21} the author described the life and times of George Adolphus Schott, who criticized the planetary atom model (\cite{Schott_39}) which was proposed by  \cite{Bohr_13} based on results of remarkable experiments done by E.~Rutherford.}. People came back to this theory in forties and fifties of the last century due to an opportunity to detect a synchrotron radiation in accelerators and in astronomy, initially, in radio band and later, in a wide spectrum of electromagnetic radiation from radio up to $\gamma$-radiation.
An existence of synchrotron radiation was re-discovered by I. Pomeranchuk\footnote{Academician I. Ya. Pomeranchuk was one of the most favorite students of L. D. Landau. Isaac  Pomeranchuk was the founder of the Theoretical Department at Institute of Theoretical and Experimental Physics (ITEP) and L. D. Landau was a half time researcher at the department. In 1998 ITEP established Pomeranchuk Prize (see, \url{https://en.wikipedia.org/wiki/Pomeranchuk_Prize}) and three Pomeranchuk Prize laureates later got a Nobel prize in physics: Yoichiro Nambu,  Roger Penrose and
 Giorgio Parisi (perhaps a number of people getting both Pomeranchuk and Nobel Prizes could be more than three but according to Pomeranchuk Prize rules of awarding a Nobel prize winner can not be chosen as a Pomeranchuk Prize laureate). L. \cite{Okun_03} wrote a short scientific biography of I.  Pomeranchuk .} and his co-authors, see papers by \cite{Pomeranchuk_40,Iwanenko_44,Artsimovich_45}.
Later, a theoretical analysis of synchrotron radiation was given by \cite{Schiff_46,Blewett_46,Schwinger_49}, the first detection
of X-ray radiation from accelerated electrons in the General Electric 70-MeV synchrotron was reported by \cite{Elder_47}. Very often paper by \cite{Schwinger_49} is treated as a key theoretical paper in the field.
 In 1950 D. Ivanenko, A. A. Sokolov and I. Pomeranchuk were awarded the State prize of the second grade for
works on synchrotron radiation, presented in book "Classical Field Theory" written by D. I. Ivanenko, A. A. Sokolov in Russian.
Famous Soviet astrophysicist I. S.  \cite{Shklovsky_46} reminded  that in forties of the last century he followed a talk about a discovery of radio emission from Sun
and he concluded that  the radio emission from Sun was generated due to a synchrotron radiation phenomenon. He also concluded that the synchrotron effect is a cause of electromagnetic radiation in wide spectral band and this idea was the most brilliant from all his ideas in his entire scientific career as it was noted in book by \cite{Shklovsky_91}.
For Crab Nebula Shklovsky interpreted electromagnetic radiation in wide spectral band (from radio to X-ray) as the synchrotron emission (\cite{Shklovsky_53,Shklovsky_76}),
see also English translation of the first edition of  book by \cite{Shklovsky_68}.
Martin \cite{Rees_71} supposed that radio emission from extended radio source  may be explained by synchro-Compton radiation (in this case electrons are accelerating by electromagnetic waves).

\cite{Shklovsky_75}
 assumed that there is a black hole at the Galactic Center with mass around $3 \times 10^4 M_\odot$
and radiation has a non-thermal origin and probably a synchrotron radiation is responsible for a significant part of radiation from the Galactic Center (earlier \cite{Linden_Bell_71} emphasized arguments supporting a necessity for a presence of supermassive black hole at the Galactic Center with mass estimates in the range $[4\times 10^3, 10^7]M_\odot$). In spite of the underestimation of the black hole mass, ideas about a presence of a black hole at Sgr A* and the synchrotron emission from the Galactic Center region received confirmation in subsequent studies.

Synchrotron radiation is a cause of energy losses in accelerators, but it plays extremely important role in astrophysics asit was noted by \cite{Ginzburg_58,Ginzburg_59,Ginzburg_65}. A more recent review on different aspects
of synchrotron radiation in astrophysical sources is given in reviews by \cite{Ginzburg_84,Ginzburg_85,Bisnovatiy_99}.

\section{\large Early VLBI in USSR}

Soviet radio engineer Leonid Matveenko was one of the first persons who understood an opportunity of inter-continental radio observations and early history of these studies is described in papers \cite{Matveenko_07a,Matveenko_07}.
In fall 1962 Matveenko reported ideas of VLBI in Pushchino at a seminar of
the Radio Astronomy Laboratory and he did not get a support to conduct such and experiment in Crimea as it was proposed.
However, these ideas were supported by participants of a seminar at Sternberg State Astronomical Institute (SSAI) of Moscow State University where
the SSAI director D. Ya. Martynov recommended to take out a patent due high scientific and technological importance of this proposal. Instead of patent
a scientific paper on the issue has been published in Soviet journal "Radiophysics" (\cite{Matveenko_65}). In this paper the authors proposed independent recording the signals and subsequent processing the data. In initial version of the paper the authors proposed to use a ground -- space interferometer but
the editorial board of the journal recommended to remove this idea from the accepted version of the paper as it was noted by \cite{Matveenko_07a,Matveenko_07}, however, we should mention also that earlier \cite{Papaleksi_47}  proposed to develop radio interferometers for geodesy and emphasized an importance of radio observations, in particular in the direction toward Galactic Center, where it was discovered a strong radio source Sgr A* (\cite{Jansky_33}).
In summer 1963  the director of the Jodrell Bank Observatory  B. Lovell visited Soviet Union as a guest of Soviet Academy of Science.   Matveenko delivered a talk about potential opportunities of interferometers with very large bases and Lovell noted that this idea looks feasible but he did not see any astronomical problem where such a resolution is needed as \cite{Matveenko_07} reminded. Both sides signed a memorandum on understanding about joint observations of Crimean and British telescopes at 32~cm wavelength. But these plans were not realized.

The VLBI technique for observations of compact bright radio sources has been proposed in USSR in sixties of the last century and these ideas were realized in the joint experiment US -- Russian experiment proposed by M. Cohen and K. I. Kellermann where 22 m Pushchino and 43 m Green Bank antennas were planned to use, but in the experiment Pushchino antenna was substituted with Simeiz one   (\cite{Kellermann_92,Lovell_73}).
Results of observations with the first Soviet -- American interferometer at 2.8~cm and 6~cm in September and October 1969 were published by \cite{Broderick_71}.
The interferometer consisted of the 42-m radio telescope of the National Astronomical Observatory at Green Bank (West Virginia) and 22-m radio telescope at Simeiz (Crimea).
Results of Soviet -- American VLBI observations were discussed in popular Soviet journal "Science and Life" where  \cite{Matveenko_73} called VLBI as "telescope with the Earth size".

In June 1971 Soviet and American teams of radio astronomers carried out interferometric observations of water vapour maser line at 1.35~cm. This interferometer was formed by the 37-m radio telescope of the Haystack Observatory at Westford (Massachusetts) \cite{Burke_72}. Several observations were done with this interferometer in 1976--1977 as it was noted by \cite{Matveenko_78}.

On April 28 and May 6, 1976 the first multi-continental interferometric observations were carried out at 1.35~cm wavelength by  \cite{Batchelor_76}.  Astronomers from four radio telescopes participated in these observations. These telescopes were the 64-m NASA telescope in Tidbinbilla (Australia), the 40-m antenna of the Owens Valley Radio Observatory (OVRO), the 26-m antenna at the Maryland Point Observatory of the US Naval Research Laboratory (NRL) and the 22-m antenna of the Crimean Astrophysical Observatory (Simeiz, Crimea).

\section{\large Projects of ground--space interferometers}

As it was noted earlier, in the first paper by \cite{Matveenko_65} on VLBI observations  the authors discussed an opportunity ground -- space observations but these sentences were removed under request of the editorial board. However, later Soviet scientists and engineers formed a working group to develop a space radio antenna to act as a space component of ground -- based interferometer. As \cite{Matveenko_07a} reminded the head of the project was V. P. Mishin, the scientific head was L. I. Matveenko, the chief engineer was V. I. Kostenko. It was assumed that the ground -- space interferometer will have an opportunity to observe compact maser sources and AGN at 1.35~cm wavelength.
Taking into account constraints of a space launcher on mass and size of the payload it was proposed to build 3.1 m radio antenna. Optimizing many parameters  of space antenna it was decided to use parabolic reflector and the Cassegrain design for irradiation as it was noted by \cite{Matveenko_82}. The orbit apogee of the spacecraft with the radio antenna was expected to be around $80\times 10^3$~km, while the orbit perigee was planned to be around $30\times 10^3$~km. Many of ideas introduced by Soviet scientists and engineers were successfully realised in the first ground -- space project VSOP (VLBI Space Observatory Programme);  HALCA (Highly Advanced Laboratory for Communications and Astronomy) or the code name MUSES-B (for the second of the Mu Space Engineering Spacecraft series) as it was noted by \cite{Matveenko_07a}. The Japanese HALCA satellite was moving along a highly elliptical orbit with an apogee altitude of 21,400 km and a perigee altitude of 560 km, with an orbital period of approximately 6.3 hours. HALCA was launched in February 1997  and made its final VSOP observations in October 2003. Initially it was planned
to observe at  in three frequency bands: 1.6 GHz, 5.0 GHz, and 22 GHz, it was found that the sensitivity of the 22 GHz band had severely degraded shape after orbital deployment and observations were done only in  1.6 GHz and 5.0 GHz. Clearly, that the highest frequency band corresponds to the best angular resolution, therefore the expected highest resolution was not reached.

In eighties the idea on a Russian space -- ground interferometer (Radioastron) started to discuss again. It was expected that the interferometer would have an angular resolution at a level of a few microarcseconds at the shortest wavelength 1.3~cm as it was noted by \cite{Kardashev_88,Kardashev_01}. However, the space antenna was launched only in 2011 and Soviet astronomers lost an opportunity to built the first ground--space VLBI radio telescope and conduct observations in 1.35~cm wavelength with the best angular resolution before the realization of Japanese HALCA mission. The Radioastron mission was successfully launched in 2011 and was operating until 2019, scientific results after five years of operation are given by \cite{Kardashev_17}.

A perspective space -- ground VLBI mission is Millimetron\footnote{https://millimetron.ru/en/general}. It will cryogenic antenna with 10 m diameter. A wavelength coverage is $[70 \mu m, 1 mm]$. Its orbit is around Lagrangian L2 point in the Sun -- Earth two body system (as many other astronomical missions). A wavelength range for  space Earth-VLBI	observations is $[0.5, 10]~$mm. A wavelength for single dish observations is $[0.07, 3]$~mm. An expected  launch date is 2029 or later.

\section{\large GR Effects in a strong gravitational field}

In seventies of the last century \cite{Bardeen_73}\footnote{A remarkable relativist James Maxwell Bardeen passed away on June 20, 2022 and
 he had an opportunity to see realizations of his theoretical picture for Sgr A* and M87*. It is a very nice case when a theoretical concept was coined in skies.} presented a picture of a dark region (a shadow) for a gedanken observation which corresponds to a bright screen located
behind a Kerr black hole and a distant observer is located in the equatorial plane. Later, \cite{Chandrasekhar_83} reproduced a similar picture in his book. However, neither Bardeen nor Chandrasekhar did not consider shadow as a possible test of GR since a) shadow sizes are extremely small to be detected for known black holes and b) there are no bright screens precisely behind black hole in astronomy. The authors represented a shadow shape as
a function $\beta(\alpha)$, where $\beta$ corresponds to impact parameter in rotation axis direction while $\alpha$ correspond to impact parameter in
in the equatorial direction.
 In addition we would like to note that perhaps  Bardeen and Chandrasekhar do not suggest to use the apparent shape of a black hole as GR test perhaps because the dark region (shadow) is too small to be detectable for all known estimates of black hole masses and distances toward them.
\cite{Falcke_00, Melia_01} simulated  shadow formation for the Galactic Center in the framework of  a toy model, where the authors took into account electron scattering for
for a radiation in mm and cm bands. The authors concluded that  it is possible to observe  a dark region (shadow) around the black hole  in mm band, while it is not possible to see a shadow in cm band due to electron scattering.
It was expected to create a global network acting in 1.3~mm wavelength, therefore the best angular resolution of this interferometer is around
 $25~\mu as$ (similar to the resolution of EHT network as it was by \cite{Akiyama_19}, while the shadow diameter was estimated as small as $30~\mu as$
 assuming that the black hole mass is $2.6 \times 10^6M_\odot$ as it was evaluated by \cite{Eckart_96,Ghez_98}), therefore, expectations for shadow observations with these facilities were not very optimistic, however, now we know that the black hole mass is more $4 \times 10^6M_\odot$
and EHT Collaboration reconstructed the shadow at Sgr A* in 2022. Therefore, a problem of shadow reconstruction using EHT Collaboration observations
 is very hard but realizable.

In times when the Radioastron mission was preparing for its launch it was known that its  best angular resolution at the shortest wavelength 1.3~cm was around 8~$\mu as$ and it was comparable with the Schwarzschild diameter for the black hole at the Galactic Center since its mass was evaluated as high as $5 \times 10^6 M_\odot$   (based on estimates done by \cite{Rees_82}). Therefore it was expected that observations with so precise angular resolution will give an opportunity to find signatures of general relativistic effects. \cite{Zakharov_05a,Zakharov_05b} proposed to use  shadow observations around the Galactic Center as a test of presence of a supermassive black hole at Sgr A* since in the case of black hole mass around  $4 \times 10^6 M_\odot$ and a distance around 8~kpc toward the Galactic Center the shadow size is around $50~\mu as$.
Usually, there are no bright screens behind astrophysical black holes, however, following ideas proposed by \cite{Holz_02} in paper by \cite{Zakharov_05a} it was noted that a shadow should be surrounded by secondary images of many astrophysical sources and a presence of these secondary sources gives an opportunity to outline a shadow. It is important to note that a presence of a shadow depends only black hole metric and
it does not depend on uncertainties of our knowledge about accretion flows and only in the case if emitting regions are very close to
black hole horizons for rapidly rotating black holes the shadow sizes and shapes may be different from the standard case of a bright screen behind a black hole.
\cite{Zakharov_05a}  showed that in the case of equatorial plane position of a distant observer the maximal impact parameter in the rotational axis direction is always (independently on $a$) $\beta_{max}(\alpha_{max})=3 \sqrt{3}$
while $(\alpha_{max})=-2a$.
This claim was based on an analysis of critical curve for Chandrasekhar parameters $\eta (\xi)$ which separates scatter and capture of photon in Kerr metric. This analysis was done by \cite{Zakharov_86}, see Fig. 2 in the paper and discussion therein and  if one considers critical values corresponding to multiple roots of the polynomial describing a radial photon motion as functions of  radial coordinate $r$  ($\xi(r)$, $\eta(r)$)and one has a maximal  $(\eta(\xi))$ at $\xi=-2a$  one has  $\eta(-2a) =27$ and $r(-2a)=3$ (see also the critical curve Fig. 34 (page 352 in book by Chandrasekhar  1983). If $\eta(\xi)$ is known, one could obtain $\beta(\alpha$). Therefore, the function  $\eta(\xi)$  determines information about shadows for any position angle.

\cite{Zakharov_05a} expressed a hope that the shadow may be detected with Radioastron facilities if electron scattering may be ignored, while the authors expressed a strong belief that the shadow can be detected with  VLBI network acting in mm band or with the projected ground--space interferometer Millimetron. The recent  results by \cite{Akiyama_22a} where the shadow was reconstructed for Sgr A* remarkably confirmed our predictions. Earlier, a shadow was reconstructed by \cite{Akiyama_19} for M87*. Later, there were presented polarization maps for M87* by \cite{Akiyama_21_a}
and possible distributions of magnetic fields were also given by  \cite{Akiyama_21_b} (polarization is connected with
synchrotron radiation of electrons accelerating in magnetic fields near M87*).

Based on results of these studies \cite{Kocherlakota_21} constrained charges of several metrics including Reissner -- Nordstr\"om,
Frolov, Kazakov -- Solodukhin and several other ones. We would like to note that blue dotted line in the left panel Fig. 2 shown by  \cite{Kocherlakota_21} corresponds to an analytical expression for the shadow size as a function of charge  and Fig. 2 as it was shown by \cite{Zakharov_05a}.

\section{\large Shadow sizes for RN black holes with a tidal charge}

\begin{figure}[th!]
\begin{center}
\includegraphics[width=0.95\textwidth]{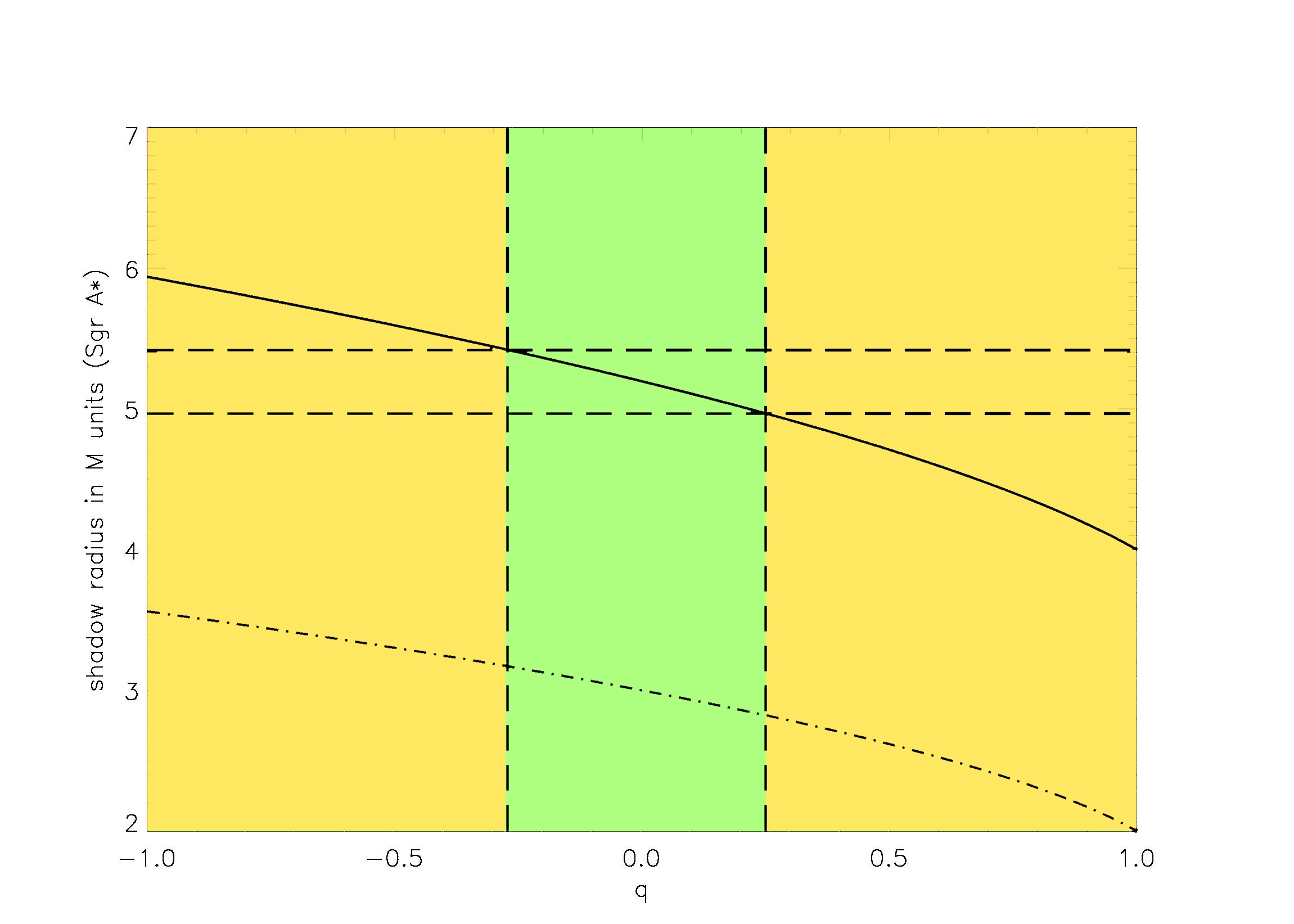}
\end{center}
\caption{Shadow (mirage) radius (solid line) and radius of the last
circular unstable photon orbit (dot-dashed line) in $M$ units as a
function of
 $q$. We adopt a shadow diameter $\theta_{\text{sh~Sgr A*}} \approx (51.8
\pm 2.3)\mu as $ at 68\% confidence levels as it was given by
\cite{Akiyama_22a}. Horizontal dashed lines correspond to constraints on shadow radius in $M$ units.
Light green vertical strip corresponds to $q$ parameter ($-0.27 < q < 0.25$) which are currently consistent with the shadow size estimate done by the EHT collaboration for Sgr A* while yellow vertical strips correspond to $q$ parameters which are not consistent with the shadow size estimate.}
 \label{Fig1}
\end{figure}

In \cite{Zakharov_05a} an analytical expression for shadow radius has been obtained as a function of a black hole charge and in the derivation we used
an algebraic condition of vanishing discriminant which was used earlier by \cite{Zakharov_91,Zakharov_94}.
However, cosmic plasma is neutral we do not expect to find significant electric charge for astrophysical black holes, but
in the framework of theories with extra dimension there are solutions which look very similar to Reissner -- Nordstr\"om ones.
For instance, \cite{Dadhich_00} showed that if we
consider the Randall--Sundrum II braneworld scenario, the Reissner -- Nordstr\"om
metric may be a black hole solution in the model. Later, this solution was called the Reissner -- Nordstr\"om 
with a tidal charge.
Similar solutions may exist in the scalar-- tensor theories which are called now Horndeski theories as it was shown by \cite{Babichev_17}.

An expression for the Reissner - Nordstr\"{o}m metric  may be written in the form in natural
units ($G=c=1$)
\begin {equation}
  ds^{2}=-\left(1-\frac{2M}{r}+\frac{\mathcal{Q}^{2}}{r^{2}}\right)dt^{2}+\left(1-\frac{2M}{r}+\frac{\mathcal{Q}^{2}}{r^{2}}\right)^{-1}dr^{2}+
r^{2}(d{\theta}^{2}+{\sin}^{2}\theta d{\phi}^{2}),
\label{RN_0}
\end {equation}
where $M$ is the mass of a black hole and $\mathcal{Q}$ is its charge (in the case of electric charge we have the ordinary Reissner -- Nordstr\"om 
metric, while  \cite{Dadhich_00} showed $\mathcal{Q}^2$ may be negative and it is called a tidal charge).

We introduce notations  $\hat {r}=r/M, \xi=L/(ME)$ and $\hat
{\mathcal{Q}}=\mathcal{Q}/M.$ Below we omit the hat symbol
for these quantities. We also introduce
 $l=\xi^{2}, q=\mathcal{Q}^{2}$.
The polynomial $R(r)$ describes  a motion along $r$-coordinate and it has a multiple root $r_{crit}$  if and only if the polynomial discriminant is vanishing and as it was shown by \cite{Zakharov_14} we found
the polynomial $R(r)$ has a multiple root for $ r\geq r_{+}$ if and only if
\begin {eqnarray}
 l^{2}(1-q)+l(-8q^{2}+36q-27)-16q^{3}=0. \label{RN_D_8}
\end {eqnarray}
 If $q=1$, then $l = 16$, or $\xi_{cr}=4$ as it was noted by \cite{Zakharov_05a,Zakharov_14}. These values also correspond to the blue curve
    shown in Fig. 2 presented
by \cite{Kocherlakota_21} since at this curve we can see that
for $\mathcal{Q}=1$ we have $\xi_{cr} (\mathcal{Q})=4$.
 From Eq. (\ref{RN_D_8}), we have
   \begin {eqnarray}
l_{\rm cr}=\frac{(8q^{2}-36q+27)+\sqrt{\left(9-8q\right)^3}}{2(1-q)}, \label{RN_D_9}
\end {eqnarray}
Therefore, we see  from the last relation that there are circular unstable
photon orbits only for $q \le \dfrac{9}{8}$.
For $1< q \le \dfrac{9}{8}$
we have naked singularities and we have unstable photon circular orbits but there are no shadows for these tidal parameters.
As it was shown by \cite{Zakharov_05a} a set of Chandrasekhar's parameters $\xi, \eta$ corresponding to the photon unstable circular orbits
separates capture and scatter regions for Kerr -- Newman black hole solutions but for generalizations of these solutions including
naked singularity cases this statement may be wrong as it was noted for naked singularities of Reissner -- Nordstr\"om metric with  $1< q \le \dfrac{9}{8}$ there are unstable circular photon orbits but shadows are not formed for these metrics.
Interesting cases of the naked singularities forming shadows were considered by \cite{Shaikh_18}.
 As it was firstly noted many years ago, the photon capture cross section for a charged
    black hole  has to be considerably smaller than the capture cross section of a
 Schwarzschild black hole as one can see in corresponding figures presented by \cite{Zakharov_05b,Zakharov_14,Kocherlakota_21}. The critical value of the impact parameter,
 characterizing the capture cross section for a Reissner -- Nordsr\"om black hole, is determined by the equation (\ref{RN_D_9}), since $\xi=\sqrt{l}$.
As it was shown by \cite{Zakharov_05b,Zakharov_14,Zakharov_22}
we can calculate the radius of the unstable
circular photon orbit (which is the same as
 the minimum periastron distance for all orbits which are scattered).
Namely,
\begin {eqnarray}
r_{\rm crit}=2\sqrt{\frac{l_{\rm cr}}{6}} \cos{\frac{\alpha}{3}},
\label{RN_D_10}
\end {eqnarray}
where
\begin {eqnarray}
\cos \alpha={-\sqrt{\frac{27}{2 l_{\rm cr}}}}. \label{RN_D_11}
\end {eqnarray}

\section{\large Constraints on a tidal charge}

Based on estimates of shadow size for M87* done by \cite{Kocherlakota_21} in paper by \cite{Zakharov_22} it  was estimated a tidal charge using
analytical expressions done \cite{Zakharov_14}, namely $q \in [-1.22, 0.814]$ at 68\% C. L. and the upper bound ($q_{upp}=0.814$) of the interval corresponds   to the upper limit $\mathcal{Q}_{upp}=\sqrt{q_{upp}} \approx  0.902$ which corresponds to the value shown by blue curve in Fig. 2 of paper by \cite{Kocherlakota_21}.

 \cite{Zakharov_22} found constraints on a tidal charge $-0.25 < q$ for Sgr A*  based on preliminary estimates of a ring width done by \cite{Lu_18}.
Below we improve an upper limit estimate for tidal charge using new EHT estimates for shadow size in Sgr A*.
Similarly to \cite{Akiyama_22a} we adopt a shadow diameter $\theta_{\text{sh~Sgr A*}} \approx (51.8
\pm 2.3)\mu as $ at 68\% confidence level. In Fig.~\ref{Fig1} we show an allowed region for a tidal charge. Horizontal dashed lines correspond to constraints on shadow radius in $M$ units. Light green vertical strip corresponds to $q$ parameter ($-0.27 < q < 0.25$) which are currently consistent with the shadow size estimate done by the EHT collaboration for Sgr A* while yellow strips correspond to $q$ parameters which are not consistent with the shadow size estimate.

\section{\large Conclusions}

We remind contributions of Russian scientists in development of synchrotron radiation and its astrophysical applications to explain spectra of
astronomical objects. We also noted Matveenko's contribution in the development of the VLBI method for astronomical observations.
Recent remarkable results of the EHT for reconstructions of shadows for black holes in Sgr A* and M87* showed a high efficiency of this technique.
As we noted there are analytical expressions for dependence of shadow sizes on (tidal) charge and this parameter can be evaluated from shadow size estimates done with VLBI observations and data analysis.

Considerations of Randall -- Sundrum theories with extra dimension were led to conclusions on the existence of Reissner -- Nordstr\"om solutions with a tidal charge. Based on our analytical expressions for shadow sizes as a function of charge we constrained  tidal charge for Sgr A* and improved our previous results done by \cite{Zakharov_22}. Similar constraints on  charge were obtained recently by \cite{Akiyama_22b}.

\subsubsection*{Acknowledgements}

The author thanks the  organizers of ICRAnet -- Isfahan Astronomy Meeting for their kind invitation to present a contribution for this activity.




\end{document}